\documentclass{emulateapj}

\def\gsim{\mathrel{\rlap{\lower 4pt \hbox{\hskip 1pt $\sim$}}\raise 1pt
\hbox {$>$}}}
\def\lsim{\mathrel{\rlap{\lower 4pt \hbox{\hskip 1pt $\sim$}}\raise 1pt
\hbox {$<$}}}

\begin{document}

\title{Young Supernovae as Experimental Sites to Study Electron Acceleration Mechanism} 

\author{
Keiichi Maeda\altaffilmark{1}
}

\altaffiltext{1}{Kavli Institute for the Physics and Mathematics of the 
Universe (Kavli-IPMU), University of Tokyo, 
5-1-5 Kashiwanoha, Kashiwa, Chiba 277-8583, Japan; 
keiichi.maeda@ipmu.jp .}

\begin{abstract}
Radio emissions from young supernovae ($\lsim $ 1 year after the explosion) show a peculiar feature in the relativistic electron population at a shock wave, where their energy distribution is steeper than typically found in supernova remnants (SNRs) and than the prediction from the standard diffusive shock acceleration (DSA) mechanism. This is especially established for a class of stripped envelope supernovae (SNe IIb/Ib/Ic) where a combination of high shock velocity and low circumstellar material (CSM) density makes it easier to derive the intrinsic energy distribution than in other classes of SNe. We suggest that this apparent discrepancy reflects the situation that the low energy electrons before accelerated by the DSA-like mechanism are responsible for the radio synchrotron emission from young SNe, and that studying young SNe sheds light on the still-unresolved electron injection problem in the acceleration theory of cosmic rays. We suggest that electron's energy distribution could be flattened toward the high energy, most likely around $100$ MeV, which marks a transition from inefficient to efficient acceleration. Identifying this feature will be a major advance in understanding the electron acceleration mechanism. We suggest two further probes: (1) {\em mm/sub-mm} observations in the first year after the explosion, and (2) X-ray observations at about 1 year and thereafter. We show that these are reachable by {\em ALMA} and {\em Chandra} for nearby SNe. 
\end{abstract}

\keywords{Acceleration of particles -- 
radiation mechanism: non-thermal -- 
shock waves -- 
supernovae: general --
supernovae: individual: SN 2011dh
}

\section{Introduction}

The most promising cosmic-ray acceleration mechanism involves a strong shock wave. In the standard DSA scenario \citep{fermi1949,blandford1978,bell1978}, particles acquire energy through multiple shock crossings between upstream and downstream. Studying two acceleration sites -- young SNe and evolved SNRs -- provides a seemingly controversial result. Young SNe expanding into CSM launch a strong shock wave. The shock wave generates/amplifies the magnetic field and accelerates electrons \citep{chevalier1998,chevalier2006b}, as is similar to SNRs. Stripped-envelope SNe, or SNe IIb/Ib/Ic, are well-studied in {\em cm} wavelengths \citep[e.g.,][]{chevalier2006b,soderberg2012,krauss2012,maeda2012a,maeda2012b, horesh2012}. They are characterized by a combination of high shock velocity and relatively low CSM density, thus providing an ideal site to study the intrinsic energy distribution of relativistic electrons, less affected by absorption or cooling than other classes of SNe \citep{chevalier1998,chevalier2006b}. The intrinsic power law index is typically found to be $p \sim 3$ (where $N(E)/dE \propto E^{-p}$) \citep{chevalier2006b}, not as efficient as predicted by the DSA mechanism \citep[$p=2$:][]{blandford1978,bell1978,ellison2000,morlino2012}. 
On the other hand, the {\em cm} emission from SNRs is generally consistent with the DSA prediction \citep[e.g.,][]{bamba2003,uchiyama2007}. 

In this paper, we suggest that this apparent discrepancy indeed provides a strong hint to understanding how electrons are injected into the DSA mechanism (\S 2). We suggest a scenario which explains the different properties of the radio-emitting electrons in a unified scheme. 
Our interpretation provides (at least) two observational signatures: (1) {\em mm} emissions in the first year after the explosion (\S 3), and (2) X-ray synchrotron emissions at about 1 years after the explosion and thereafter (\S 4). The paper is closed in \S 5 with concluding remarks. 

\begin{figure*}
\hspace{-2cm}
        \begin{minipage}[]{0.95\textwidth}
                \epsscale{1.5}
                \plotone{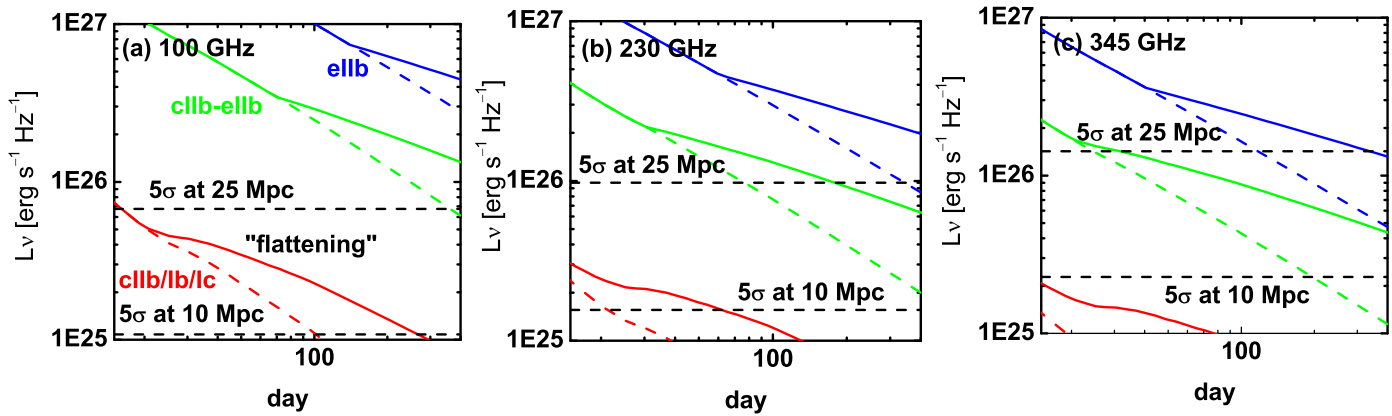}
        \end{minipage}
\vspace{-1.5cm}
\caption
{Examples of multi frequency light curves of SN-CSM interaction. Our reference model (`cIIb/Ib/Ic' shown in red) is the one which fits to the spectra and light curves of SN cIIb 2011dh in {\em cm} wavelengths, where $A_{*} = 4$. For the same model, the one with the spectral flattening at $\gamma = 200$ and without it are shown by solid and dashed, respectively. We also examine models where only the CSM density is changed, to $A_{*} = 200$ (`eIIb') and $A_{*} = 50$ (`cIIb-eIIb'). The $5\sigma$ detection limits in the continuum observation mode of {\em ALMA} with an exposure of one hour in each Band 3, 6, 7 are shown by dashed lines. 
\label{fig1}}
\end{figure*}

\section{Electron Acceleration toward The DSA Mechanism}

A major issue remains on the acceleration of electrons. For the standard DSA mechanism \citep{fermi1949,blandford1978,bell1978} to work effectively, the particles must already have enough kinetic energy (at least $\gamma \gsim 200$ for electrons in SNe IIb/Ib/Ic, where $\gamma$ is the Lorentz factor; see below). Numerical simulations predict that the energy distribution in the  high energy regime follows the linear theory of DSA (with $p \sim 2$) while in the lower energy it can be steeper \citep{ellison2000,morlino2012}. First, let us provide a simple argument on at which energy the electrons have a sufficient energy to be accelerated by the efficient DSA-like mechanism. The width of the collisionless shock wave can be approximated by the gyro radius of downstream {\em thermal} protons, while the mean free path of relativistic electrons can be approximated by the gyro radius of the {\em relativistic} electrons. Then, the condition that the electron's mean free path exceeds the shock wave width is expressed as 
\begin{equation}
E \gsim m_{\rm e} c^2 \left(\frac{m_{\rm p}}{m_{\rm e}}\right) \left(\frac{V}{c}\right) \sim 100 \ {\rm MeV} \ ({\rm i.e., } \  \gamma \gsim 200) \ ,
\end{equation}
where $V \sim 0.1 c$ is the shock wave velocity typically seen in young SNe IIb/Ib/Ic. The same criterion is $\gamma \gsim 20$ for SNRs with $V \sim 0.01c$. This assumes the Bohm limit for the strength of random magnetic field turbulence which is likely realized in the SN-CSM interaction, but if it is not the case the above criterion will go down. Above this energy, electrons can experience the whole compression ($r = 4$), and therefore $p = (r+2)/(r-1) = 2$ in the test particle limit. On the other hand, the acceleration of electrons below this energy can be totally different: Either the non-DSA pre-acceleration mechanism dominates, or these electrons are partly accelerated by the DSA in the non-linear regime \citep{ellison2000} where the subshock compression ratio is reduced by the feedback of the proton acceleration \citep{tatischeff2009}. These processes expected for these low energy electrons could result in a steep distribution with $p > 2$.

Now, we compare this critical energy scale to the energy of radio-synchrotron emitting electrons  (in {\em cm} wavelengths). The electron's energy and the corresponding synchrotron emission frequency are connected by 
\begin{equation}
\gamma \sim 80 \nu_{10}^{0.5} B^{-0.5} \ ,
\end{equation}
where $\nu \equiv 10^{10} \nu_{10}$ Hz, and $B$ is the magnetic field strength behind the shock in gauss. Observationally, $B \sim 1$ G in SNe IIb/Ib/Ic and $B \sim 100 \mu$ G in evolved SNRs. We note this is consistent with an expectation from equipartition: In equipartition ($B^{2}/4\pi \equiv \epsilon_{B} \rho_{\rm CSM} V^2$) and for the same reference value of $\epsilon_{B} \sim 0.1$, we expect $B \sim 1$ G for young SNe ($n_{\rm CSM} \sim 10^6$ cm$^{-3}$ and $V \sim 0.1 c$) and $B \sim 100 \mu$ G for SNRs ($n_{\rm ISM} \sim 1$ cm$^{-1}$ and $V \sim 0.01 c$). At 10 GHz typical of {\em cm} wavelengths, electrons' energy responsible to the emission is $\gamma \sim 80$ in SNe IIb/Ib/Ic and $\gamma \sim 8000$ in SNRs. 

Following the above estimate, it is clear that the electrons emitting in {\em cm} wavelengths are totally in different energy regimes in young SNe and SNRs. The electrons' energy responsible for the radio emission is certainly much above the DSA requirement for SNRs, so they are likely efficiently accelerated. On the other hand, the radio-emitting electrons in young SNe unlikely satisfy the efficient DSA condition, i.e., the electrons do not feel the shock wave as infinitesimal discontinuity. We suggest this is a reason why different intrinsic slopes are obtained for young SNe and SNRs. 

This interpretation offers a unique opportunity to study the electron acceleration mechanism, or the injection problem, through the SN-CSM interaction signals from young SNe. Our scenario predicts that the flattening of the intrinsic electron spectrum should take place above the energy currently probed in {\em cm} wavelengths. We propose two methods to further probe this issue using currently operating observatories: (1) {\em mm} observations (e.g., {\em ALMA}) and (2) X-ray observations (e.g., {\em Chandra}). 

\begin{figure*}
\hspace{-1.5cm}
        \begin{minipage}[]{0.95\textwidth}
                \epsscale{1.4}
                \plotone{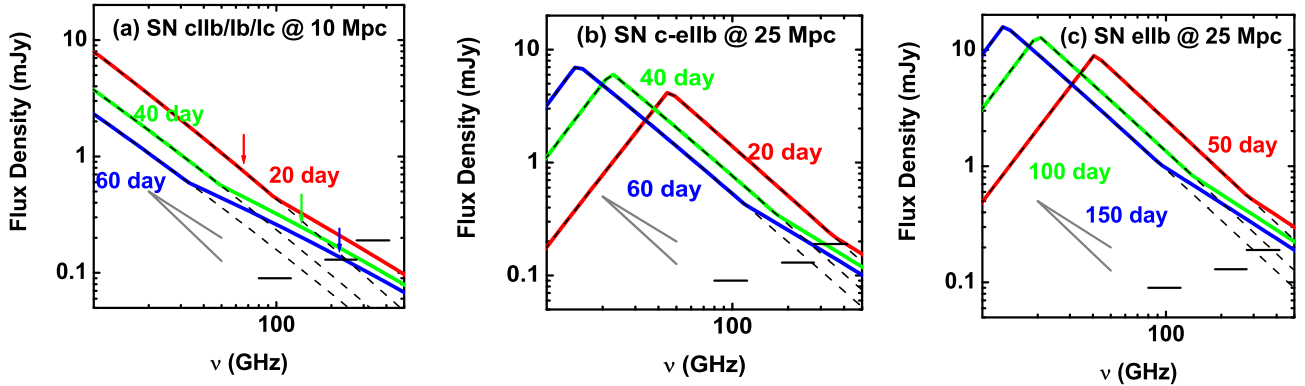}
        \end{minipage}
\vspace{-1.5cm}
\caption
{The spectral evolution of the model shown in Figure 1 (with $\gamma_{\rm fl} = 200$) at different epochs. The characteristic synchrotron cooling frequency is indicated by arrows for each snap shot for the model with $A_{*} = 4$ (`cIIb/Ib/Ic'). For the other models, the cooling frequency is below the frequency range shown here. In each snapshot, a dashed curve is for the model without flattening. On the left-bottom corner, the power law behaviors are illustrated for the cooling-dominated synchrotron spectrum with $p = 3$ (derived in {\em cm} observations for a majority of SNe cIIb/Ib/Ic) and $p = 2$ (expected above the spectral flattening energy). The {\em ALMA} $5\sigma$ continuum sensitivities with an exposure of one hour are shown by black solid line in each band. 
\label{fig2}}
\end{figure*}

\section{{\em mm/sub-mm} Emission (e.g., {\em ALMA})}

At 100 GHz, the corresponding Lorentz factor of the synchrotron emitting electrons is $\gamma \sim 250$ for $B \sim 1$ G, which is the energy range in which the flattening could already happen announcing the possible transition from inefficient to efficient electron acceleration (\S 2). Figures 1 and 2 show an example of expected multi-band light curves in the {\em ALMA} bands, and the predicted spectral evolution for the same models. The model has been constructed through standard formalisms \citep{fransson1998,bjornsson2004}, and fits well to the multi band radio light curves of nearby SN IIb 2011dh in $1.4 - 36$ GHz \citep[see][for the model details]{maeda2012a}. Here, we include the effect of the spectral flattening (at $\gamma_{\rm fl}$), while fixing all the other parameters as calibrated for SN 2011dh. We fix $\gamma_{\rm fl} = 200$, and $p=3$ and $p=2$ below and above $\gamma_{\rm fl}$. The inclusion of the spectral flattening does not affect the properties in {\em cm} wavelengths. 

This model is shown in the figure as `cIIb/Ib/Ic'. We assume a steady-state wind for the CSM density as $\rho_{\rm CSM} = 5 \times 10^{11} A_{*} r^{-2}$ g cm$^{-3}$ ($r$ is the distance from the progenitor in {\em cm}), and we adopt $A_{*} = 4$ for SN IIb 2011dh \citep{soderberg2012,maeda2012a}. The stripped-envelope SNe (IIb/Ib/Ic) are characterized by the absence/weakness of lines from particular elements in the optical maximum-light spectra \citep{filippenko1993,filippenko1997}: Weak H lines in SNe IIb, (generally) absence of H lines in SNe Ib, and neither H nor He in SNe Ic. The origin of this difference is generally interpreted as a different amount of envelope stripped off during the hydrostatic evolutionary phase \citep{nomoto1993,woosley1994}. SNe IIb are (sometimes) further sub-divided into SNe eIIb (`extended') and SNe cIIb (`compact') \citep{chevalier2010}. SNe eIIb, cIIb, Ib/Ic are then linked to the different size of the progenitor envelope, where SNe eIIb are believed to come from a Red Supergiant (RSG) while SNe Ib/c are from a compact Wolf-Rayet (WR) \citep{chevalier2010}. Their different properties in radio wavelengths can be interpreted as mainly controlled by different CSM density rather than the SN properties \citep{maeda2012b}\footnote{Peak radio date and luminosity of SNe IIb/Ib/Ic follow the expectation in which CSM density is varied while the similar explosion properties are applied \citep{maeda2012b}.}, where the CSM is as dense as $A_{*} \sim 200$ for SNe eIIb \citep{fransson1996,fransson1998}, while it is less dense in SNe cIIb/Ib/Ic ($A_{*} \lsim 10$) \citep{chevalier2006b}. To take into account this variation, we also investigate SNe with more dense CSM, named `eIIb' ($A_{*} = 200$) and a transitional case `cIIb-eIIb' ($A_{*} = 50$). Note that even with $A_{*} = 200$, the free-free absorption is estimated to be negligible at $\sim 100$ GHz unlike in the {\em cm} wavelengths \citep{chevalier2006a}, highlighting another useful nature of the high frequency observation -- the transparency even for SNe within moderately high-density environments (e.g., SNe eIIb, IIp). 

The spectral flattening in the electron energy distribution results in the flattening in the synchrotron spectrum. In light curves this is seen as a flattening in the decay rate, which takes place first in the higher frequency then eventually in the lower frequency.  For denser CSM the transition is further delayed and the large luminosity is attained due to the higher magnetic field content and higher relativistic electron density for denser CSM. 

The model curves are compared to the $5\sigma$ detection limit in the continuum observation mode of {\em ALMA} with one-hour exposure in each band \citep{brown2004}.\footnote{We have used the Cycle 1 ALMA Observing Tool for the sensitivity calculation (http://www.almaobservatory.org/).}  Figure 1 shows that {\em ALMA} can detect the synchrotron signal from nearby SNe, and that 
the possible spectral fattening can be investigated. If $\gamma_{\rm fl} \sim 200$ as in our reference value (\S 2), this will be clearly identified in $100 - 345$ GHz. For the Cycle-1 {\em ALMA} and an exposure of one hour, this is reachable for typical SNe cIIb/Ib/Ic up to $\sim 10$ Mpc, and to $\sim 25$ Mpc or even more for SNe eIIb and the intermediate case. For the larger value of $\gamma_{\rm fl}$, the beginning of this light curve flattening is delayed at the lower flux level. 

Figure 2 shows the evolution of the synthesized spectrum as a function of time, for $\gamma_{\rm fl} = 200$. The spectral break frequency (corresponding to $\gamma_{\rm fl}$) moves toward the lower frequency as time goes by. The break appears within the {\em ALMA} bands at $\sim 20 - 100$ day, and later enters into the lower frequency bands. Thus, the multi-epoch follow-up is crucial, and coordinated follow up in {\em cm} wavelengths is strongly encouraged. 

In these models, the synchrotron cooling is the dominant process in the energy range of interest here \citep{maeda2012a}. The cooling frequency evolves toward the higher energy. For the `cIIb' model, initially it is below the {\em ALMA} bands, and later moves into the {\em ALMA} bands and even to the higher energy. This evolution is in the opposite direction to the one caused by the intrinsic spectral flattening, thus these two effects are distinguishable. It will give an independent estimate of the magnetic field strength through the relation $t \sim 110 \nu_{10}^{-0.5} B^{-1.5}$ days, providing a rare case to test different scenarios for magnetic field generation/amplification. 

\begin{figure}
\hspace{-0.5cm}
        \begin{minipage}[]{0.45\textwidth}
                \epsscale{1.35}
                \plotone{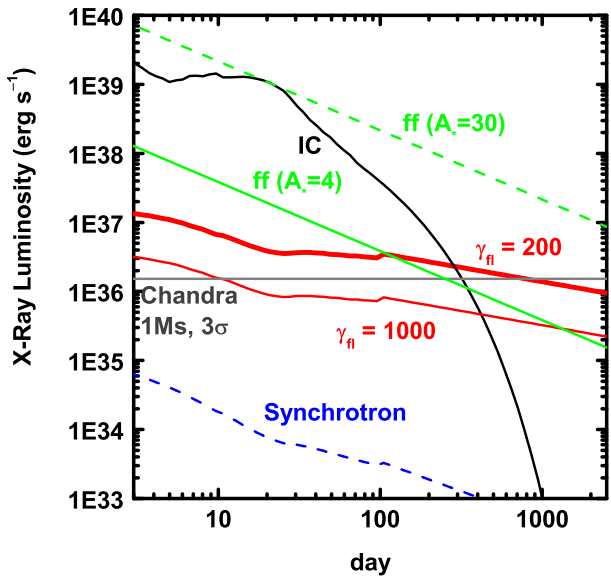}
        \end{minipage}
\hspace{-1cm}
\caption
{Predicted long-term X-ray light curve of SN 2011dh, including the effect of the possible spectral flattening. The model is the same with `cIIb' model ($A_{*} = 4$) adopted from \citet{maeda2012a}. The IC (black) is from the same model which fits the observed X-ray light curve by the IC \citep{maeda2012a}. The contribution from the free-free emission is shown by green, where the one with $A_{*} = 4$ (solid) is our reference model while the other with $A_{*} = 30$ (dashed) is the high density CSM model to account for the early X-ray emission by the free-free emission. The `original' synchrotron X-ray emission is shown by blue dashed line. The synchrotron X-ray emission with the spectral flattening is shown by red solid lines, with $\gamma_{\rm fl} = 200$ (thick) and $\gamma_{\rm fl} = 1000$ (thin). At $\sim 10 - 100$ days, the synchrotron X-ray light curve shows slight suppression from the expected power law behavior due to an additional IC cooling. The $3\sigma$ detection limit of Chandra X-ray Observatory with an exposure of 1 Ms is shown by gray: Here, we scaled the detection limit obtained for SN 2011fe \citep{margutti2012} to the distance to SN 2011dh and to the exposure of 1 Ms (assuming the Poisson-noise dominated). 
\label{fig3}}
\end{figure}

\section{X-ray Emission (e.g., {\em Chandra})}

Another suggestion for the spectral flattening is to look into a late-time X-ray property \citep[see also][]{chevalier2006b}. Figure 3 shows an expected X-ray light curve evolution ($\nu L_{\nu}$ at $\sim 1$ keV). This model assumes an additional low energy relativistic electron population at $\gamma \lsim 50$ to account for the X-ray luminosity at $\lsim 40$ day by the Inverse Compton (IC) scattering, together with the power-law synchrotron-emitting electrons at $\gamma \gsim 50$. For the IC, we assume that the SN bolometric light curve follows the $^{56}$Co decay after 100 days. This would overestimate the IC scattering effect there, so the IC contribution could decreases more quickly than shown in Figure 3 after 100 days. In addition to this reference model, \citet{maeda2012a} provided another possibility, where the CSM density is as high as $A_{*} \sim 30$ and the X-ray is dominated by the free-free emission from the thermal electrons \citep[see also][]{campana2012,sasaki2012}. While \citet{maeda2012a} \citep[see also][]{soderberg2012} preferred the low-density solution in view of a connection to other SNe Ib/c, the early-phase observation of SN 2011dh itself did not discriminate these two possibilities. 

The synchrotron contribution is small as compared to other mechanisms if we extrapolate the intrinsic electron energy distribution derived for the radio emitting electrons ($\gamma \sim 100$) to the X-ray synchrotron emitting electrons (i.e., without the flattening). With the spectral flattening the synchrotron X-ray is enhanced, and can dominate the free-free emission at $\gsim 100$ days if $\gamma_{\rm fl} \sim 200$ (if $A_{*} = 4$). It further dominates the IC X-ray at $\gsim 300$ days. On the other hand, the free-free always dominates the X-ray emission if the CSM density is as high as $A_{*} = 30$. This provides a test for the origin of the X-ray emission from SN 2011dh. If the early-phase X-ray was due to the high density CSM and the free-free emission, then the late-phase X-ray luminosity should stay as luminous as $10^{37}$ erg s$^{-1}$ even at a few years after the explosion. Otherwise, it should be below $\sim 10^{36}$ erg s$^{-1}$. 

The free-free emission follows the temporal evolution of $\nu L_{\rm ff} \propto t^{-1}$ \citep{chevalier2006a,chevalier2006b}. The synchrotron X-ray is in the cooling regime, and roughly follows $\nu L_{\rm syn} \propto t^{-0.8}$ without the flattening \citep{chevalier2006b,maeda2012b}. With the flattening at $\gamma_{\rm fl} = 200$, the synchrotron X-ray emitting electrons are above this energy, and thus $\nu L_{\rm syn} \propto t^{-0.3}$ \citep{chevalier2006b,maeda2012b}. 

In case that the early-phase X-ray was dominated by the IC (which is testable from the late-phase observation as mentioned above), it provides an interesting possibility to probe the electron acceleration mechanism \citep{chevalier2006b}. If there is the spectral flattening, the synchrotron X-ray emission can dominate in the late-phases, and this effect can be identified by the characteristic relatively flat evolution with nearly a constant luminosity ($\propto t^{-0.3}$). If $\gamma_{\rm fl} = 200$, this behavior can be detectable by {\em Chandra} {\em now} (Fig. 3). 

\section{Concluding Remarks}

In this paper, we have proposed new approaches to investigate the injection and acceleration problem of relativistic electrons at a strong shock wave. A peculiar feature derived from synchrotron radio emissions (in {\em cm} wavelengths) from nearby young SNe, especially established for stripped-envelope SNe (IIb/Ib/Ic), is the steepness in the intrinsic energy distribution of the radio-emitting relativistic electrons. We suggest this steep distribution could result from a pre-acceleration mechanism able to energize electrons up to the energy threshold needed to start the DSA, as the {\em cm} emission is produced by electrons whose mean free path is smaller than the shock wave width under the physical conditions realized in young SNe. 

According to this interpretation, we suggest that the relativistic electron's energy distribution could be flattened toward the high energy. We provide an estimate about the energy scale of electrons as $\sim 100$ MeV (i.e., $\gamma_{\rm fl} \sim 200$) where this could most likely happen. We suggest that identifying (or constraining) this feature can provide an important clue for the still-unresolved electron injection problem, and propose two practically accessible diagnostics on this issue, (1) {\em mm/sub-mm} observations in the first one year, and (2) late-time X-ray observations around one year after the explosion or thereafter. The {\em mm/sub-mm} emission from stripped-envelope SNe is detectable up to $\sim 25$ Mpc by {\em ALMA}. The second is applicable up to $\sim 10$ Mpc by {\em Chandra}, and currently testable for SN IIb 2011dh in M51. 

Besides investigating the possible flattening of relativistic electrons' distribution, these methods will provide additional information. The {\em mm/sub-mm} observation can be used to place an independent constraint on the magnetic field strength behind the shock wave. The late-time X-ray observation is able to distinguish the X-ray production mechanism in the early phase. Also, we emphasize the additional uniqueness of the {\em mm/sub-mm} observation - the transparency even for SNe eIIb and possibly SNe IIp within dense CSM. By combining the {\em cm} and {\em mm} observations, both optically thick and thin emissions can be traced to derive physical quantities behind the shock wave \citep[see, e.g.,][]{maeda2012a}. We also note that this transparency increases the applicability of the method to study the shock breakout using the early-phase synchrotron emission \citep{maeda2012b}, to SNe with more dense CSM than in {\em cm}. 

Our model prediction is based on the model which explains the {\em cm} emission from nearby SN cIIb 2011dh \citep{maeda2012a}. SN 2011dh is typical among SNe cIIb/Ib/Ic in its radio properties \citep{soderberg2012}, and there is a wide distribution of SNe IIb/Ib/Ic in the synchrotron luminosity \citep{chevalier2006a,chevalier2010}. There are radio strong SNe IIb/Ib/Ic, more luminous than SN 2011dh by an order of magnitude or even more, including `engine-driven'  SNe sometimes associated with GRBs \citep{soderberg2010}. For these especially radio strong SNe, the {\em ALMA} detectable horizon extends even to the greater distance, i.e., $\sim 100$ Mpc.

\acknowledgements 
This research is supported by World Premier International Research Center
Initiative (WPI Initiative), MEXT, Japan. K. M. acknowledges financial support by Grant-in-Aid for Scientific Research for young Scientists (23740141).

\end{document}